\documentclass[]{aa} % referee
\usepackage{epsfig}
\usepackage[varg]{txfonts}

\title{Direct deconvolution of radio synthesis images using $L_1$ minimisation}
\author{Stephen J. Hardy}
\institute{POBox 6, Pymble BC, 2073 Australia. \email{sjh@pobox.com}}

\date{Recieved 05/05/2013 / Accepted 12/08/2013}

\begin{document}

\abstract{}
{
%aims
We introduce an algorithm for the deconvolution of radio synthesis images that accounts for the non-coplanar-baseline effect, allows multiscale reconstruction onto arbitrarily positioned pixel grids, and allows the antenna elements to have direcitonal dependent gains.
}
{
%methods
Using numerical $L_1$-minimisation techniques established in the application of compressive sensing to radio astronomy, we directly solve the deconvolution equation using GPU (graphics processing unit) hardware. This approach relies on an analytic expression for the contribution of a pixel in the image to the observed visibilities, and the well-known expression for Dirac delta function pixels is used along with two new approximations for Gaussian pixels, which allow for multi-scale deconvolution. The algorithm is similar to the CLEAN algorithm in that it fits the reconstructed pixels in the image to the observed visibilities while minimising the total flux; however, unlike CLEAN, it operates on the ungridded visibilities, enforces positivity, and has guaranteed global convergence. The pixels in the image can be arbitrarily distributed and arbitrary gains between each pixel and each antenna element can also be specified.
}
{ 
%results
Direct deconvolution of the observed visibilities is shown to be feasible for several deconvolution problems, including a 1 megapixel wide-field image with over 400,000 visibilities. Correctness of the algorithm is shown using synthetic data, and the algorithm shows good image reconstruction performance for wide field images and requires no regridding of visibilities. Though this algorithm requires significantly more computation than methods based on the CLEAN algorithm, we demonstrate that it is trivially parallelisable across multiple GPUs and potentially can be scaled to GPU clusters. We also demonstrate that a significant speed up is possible through the use of multi-scale analysis using Gaussian pixels.
}
{
}


\maketitle
\section{Introduction}

There has been significant interest and activity in new algorithms for the synthesis of images from radio interferometric measurements over the last 10 years, driven by the conceptualisation, design and partial commissioning of several new radio observatories that have significantly different scale and coverage than those in the past. New facilities such as the Australian Square Kilometre Array Precursor\footnote{http://www.atnf.csiro.au/projects/mira/} (ASKAP), the Low Frequency Array\footnote{http://www.lofar.org/}, the Murchison Widefield Array\footnote{http://www.mwatelescope.org} (MWA) all share the characteristic of generating huge amounts of observational data, as well as each having unique challenges due to their individual design choices. 

One of the basic challenges shared by each of these facilities is the non-coplanar baseline effect. This effect is present for any radio interferometer that has baselines that are not aligned in the E-W direction. For these baselines, the rotation of the earth moves the baselines of the telescopes into planes that are tilted with respect to their initial orientation. This  introduces a component of the baseline in the direction of the source (the $w$-component) that leads to a defocus effect that must be compensated for during image processing. This effect becomes more important for wider fields of view, longer baselines and lower frequencies (due to the wider field of view). There are a variety of established methods for dealing with the non-coplanar base effect, for instance, the W-projection algorithm (\cite{Cornwell:2005p1977}) and W-snapshots (\cite{Cornwell:2012p1785}). 

Another important effect, particularly for the low frequency instruments, is that of direction dependent gains. In most radio telescopes, the gain of the antenna is largely a function of the direction angles relative to the pointing direction. This is described by the pattern of the primary beam, $A(l,m)$, where $l$ and $m$ are direction cosines relative to the pointing direction. However, there is also some dependence on the absolute pointing direction of the telescope relative to the ground, making the pattern of the primary beam a function of the zenith angle, $Z$, and parallactic angle, $\chi$, as well. For electronically steered low frequency telescopes such as the MWA, highly accurate compensation for this effect is a key issue.

In this paper we introduce an algorithm for synthesis image deconvolution called SL1M (Synthesis through $L_1$ Minimisation), that can deal with arbitrary collections of non-zero coplanar baselines and direction dependent gains.

Three major approaches to image deconvolution in radio synthesis astronomy have been taken in the past. By far the most prevalent of these is the family of algorithms based on the work of \cite{Hogbom:1974p2738} called CLEAN. For this family of algorithms, it is assumed that the image can be represented as a small set of sources, either points sources in the original approaches, or extended sources as is the case in multi-scale CLEAN (\cite{Cornwell:2008p1783}).  For the classic CLEAN algorithm, the image is reconstructed by iteratively determining the point source that best fits the observed visibilities and adding some fraction of best fit flux from that source to the image. This process is repeated until some convergence requirement is met. It was shown by \cite{Marsh:1987uc} that for sufficiently separated point like sources, the CLEAN algorithm is equivalent to solving the deconvolution problem by fitting the observed visibilities while minimising the total reconstructed flux intensity (that is, the sum of the pixels intensities of the image). The SL1M algorithm represents an alternate direct method for fitting the observed visibilities while minimising the total reconstructed flux intensity.

A second set of algorithms is based on the constraint that the entropy should be maximised, for example in \cite{Narayan:1986p3297}. These algorithms are of less relevance to this work and will not be discussed further.

A third, more recent, set of algorithms is based around the ideas of compressive sampling (CS). Compressive sampling was introduced to radio synthesis astronomy in \cite{Wiaux:2009p3076}, where the Basis Pursuit algorithm was applied to reconstruct images from visibilities with coplanar base-lines. This work was extended in \cite{Wiaux:2009p3050} to the case of baselines that had a constant non-coplanar component and that demonstrated how this component introduced a spread spectrum effect that improved the Basis Pursuit reconstruction. This was further extended in \cite{McEwen:2011p3034} where the Basis Pursuit reconstructions were performed for a wide-field on a non-rectangular grid, but still under the constraint of a constant $w$ for all baselines. Most recently, \cite{Carrillo:2012ho} have introduced the SARA (Sparsity Averaging Reweighted Analysis) algorithm that optimises the data fit while regularising with respect to the average signal sparsity simultaneously in multiple wavelet bases. 
 
The SL1M algorithm solves a similar problem to the CS reconstruction problem introduced by \cite{Li:2011p1778}.  In \cite{Li:2011p1778}, it is assumed that the image is sparse (few non-zero components) in some basis, and $L_1$ minimisation is used to determine the image that best agrees with the observed visibilities and has the minimum $L_1$ norm in the selected basis. Their technique was demonstrated for the Dirac basis and for the isotropic undecimated wavelet basis and showed image quality improvements over reconstruction with the CLEAN algorithm. Note that \cite{Carrillo:2012ho} demonstrated substantially better reconstruction performance using SARA compared to reconstructions with the isotropic undecimated wavelet transform. \cite{Wenger:2010p2327} have also explored solutions to the sparse reconstruction problem based on total flux minimisation, and demonstrated improvements over the CLEAN algorithm in a Daubechies wavelet basis. We make a brief comparison of the theoretical basis of this work and these other approaches in Section \ref{sec:compare}.

Rather than operate on gridded visibilities and use the Fourier transform to transform between the visibility domain and the image domain, as is done in \cite{Li:2011p1778}, the SL1M algorithm works with raw visibilities and uses the full matrix transformation between visibility space and image space to switch domains. While this method is highly computational, it has some benefits in terms of flexibility - in particular it can model direction dependent gains explicitly, naturally deals with non-coplanar baselines, and also allows sampling on non-rectangular grids. It is also based on $L_1$ minimisation, and uses the same $L_1$ minimisation algorithm as used in \cite{Li:2011p1778}. 

To describe this method and its relationship to existing algorithms, we first make a brief introduction to the deconvolution problem in radio synthesis imaging and then outline the basic approach taken in SL1M for solving the problem for point source and Gaussian pixels. We then describe the implementation details, particularly the parallelisation strategy necessary to make the algorithm computable in a reasonable amount of time. Next, we apply SL1M to some simple simulated datasets to illustrate the features and constraints of the approach, and then apply it to two real datasets which demonstrate the efficacy of the algorithm. In the following section we demonstrate a version of the algorithm with improved algorithmic performance based on multi-scale analysis using the Gaussian pixel basis. After this we make a theoretical comparison to existing work, followed by concluding remarks and possible future avenues of investigation.

\section{Defining the direct solution to the deconvolution problem} \label{sec:direct}

To begin, we define the coordinate systems for the problem. Consider the visibilities measured by a two element radio interferometer with baseline $\bf b$, pointing at the sky in a direction ${\bf s}_0$. The baseline, $\bf b$, can be represented in terms of rectilinear coordinates $(u,v,w)$, so that ${\bf b} = \lambda( u{\bf e}_u + v{\bf e}_v+w{\bf e}_w )$, where the orthonormal basis vectors $({\bf e}_u,{\bf e}_v,{\bf e}_w)$ are defined such that ${\bf e}_w={\bf s}_0$ and ${\bf e}_u$ and ${\bf e}_v$ are aligned with a convenient axes, such as East and North. Sky coordinates, $(l,m,n)$, are defined where $l$ and $m$ are parallel to $u$ and $v$ respectively and $n$ is parallel to $w$. As the sky coordinates are restricted to the celestial sphere, $n=\sqrt{1-l^2-m^2}$.

Given some brightness distribution on the sky, $I(l,m)$, and a receptive pattern of the primary beam, $A(l,m;Z,\chi)$, the spatial coherence of the radiation field observed by an interferometer (the visibilities) with a baseline represented by $(u,v,w)$ can be expressed as 

\begin{equation}
V \left( u,v,w \right) = \int{ \frac{A(l,m;Z,\chi)I(l,m)}{\sqrt{1-l^2-m^2}} e^{-2 \pi i  \left(ul+vm+w\left(\sqrt{1-l^2-m^2}-1\right)\right)} dl\,dm}. \label{eq:fullvis}
\end{equation}

Note that (\ref{eq:fullvis}) is also a function of observed frequency and polarisation, in that the visibilities are generally measured at many different frequencies, and in different polarisations. Dependence on frequency and polarisation will not be described here - the algorithms for deconvolution can be applied to either line or continuum channels and each polarisation independently.

When the relation $(l,m)\ll 1$ holds, then (\ref{eq:fullvis}) reduces to a Fourier transform of the sky brightness distribution multiplied by the primary beam (dropping the direction-dependence), as given by
\begin{equation}
V \left( u,v \right) = \int{ A(l,m) I(l,m)e^{-2 \pi i \left(ul+vm\right)} dl\, dm} \label{eq:ftvis}
\end{equation}
and all dependence on the $w$ factor is lost in the relationship. However, as noted above, the $w$ term for many observations is significant, and neglecting it can lead to artefacts and inaccuracies in the the deconvolved image.

Examining equation (\ref{eq:ftvis}) is instructive as it highlights the basic problem of radio synthesis imaging. To reconstruct an image $I(l,m)$ to a given resolution, it is necessary to know all the visibilities in the plane $(u,v)$ out to the Nyquist frequency of the image that is to be reconstructed. However, only a fraction of the visibilities are observed, and so the inverse problem for equation (\ref{eq:ftvis}) is under-constrained. This under-constrained problem can then only be solved by introducing new constraints, based on a-priori knowledge or assumptions about the properties of the image.

To proceed, we discretise the visibility equation (\ref{eq:fullvis}). Firstly, if the $N_v$ observed visibilities are written as $V_j(u_j,v_j)$, then the relation between the measured visibilities and the observed image is

 \begin{equation}
\begin{array}{ll}
& V_j\left( u_j,v_j,w_j \right) = \\
& \hspace{0.5cm}  \int{ \frac{A(l,m;Z_j,\chi_j)I(l,m)}{\sqrt{1-l^2-m^2}}e^{-2 \pi i  \left(u_j l+v_j m+w_j \left(\sqrt{1-l^2-m^2}-1\right)\right)} dl\,dm}.   \label{eq:samuvw}
\end{array}
\end{equation}

where the dependence of the zenith and parallactic angles on the visibility being observed has been included (as different sets of visibilities will be observed at different times, and hence at different angles on the sky).

Modelling the image as a sum of functions, $f_k(l,m)$, then we may write equation (\ref{eq:samuvw}) as 

 \begin{equation}
\begin{array}{ll}
&V_j\left( u_j,v_j,w_j \right)  =  \\ 
& \hspace{0.5cm} \sum_{k}\int{ \frac{A(l,m;Z_j,\chi_j)f_k(l,m)}{\sqrt{1-l^2-m^2}}e^{-2 \pi i  \left(u_j l+v_j m+w_j \left(\sqrt{1-l^2-m^2}-1\right)\right)}} dl\,dm,   
\end{array}
\label{eq:samlm}
\end{equation}
and the relationship between the visibilities and the image can be evaluated for different classes of functions.

\subsection{Delta function pixels}

As a first approach, the sky brightness is modelled as a weighted sum of delta functions. To facilitate changing between a 2 dimensional image coordinate system and a 1 dimensional image coordinate systems (for the use of linear algebra), we introduce a list of two dimensional coordinates $(l_k, m_k)$ indexed by a linear index $k$ which enumerates each pixel being modelled.

As a simple example, $(l_k, m_k)$ may describe an $N_l$ by $N_m$ grid of sample points. The $l_k$ and $m_k$ are  integer coordinates ranging from $-N_l/2$ to $N_l/2-1$, and the relationship with linear index $k$, which ranges from $0$ to $N_l N_m -1$ is given by
\begin{equation}
	l_k = \left(k  \bmod{N_l} - N_l/2\right) \Delta
\end{equation}
\begin{equation}
	m_k = \left(\lfloor k  / N_m \rfloor - N_m/2 \right) \Delta
\end{equation}
where $\Delta$ is the grid spacing in sine coordinates. Note that nothing in the following requires that the pixels be placed on a grid, hence the SL1M algorithm may be used for irregularly distributed pixels.

Using this approach, each function contributing to the image may be written as $f_k(l,m) = I_{l_k m_k} \delta(l-l_k)\delta(m-m_k)$, and equation ({\ref{eq:samlm}}) becomes
 \begin{equation}
\begin{array}{ll}
& V_j\left( u_j,v_j,w_j \right) =\\
& \hspace{0.5cm}  \sum_{k}\frac{A_{l_k m_k}^{(j)}}{\sqrt{1-l_k^2-m_k^2}}e^{-2 \pi i  \left( u_j  l_k+ v_j m_k+w_j \left(\sqrt{1-l_k^2-m_k^2}-1\right)\right)}  I_{l_k m_k} .   
\end{array}\label{eq:samdelta}
\end{equation}
where $A_{l_k m_k}^{(j)} = A(l_k,m_k;Z_j,\chi_j)$.

Equation (\ref{eq:samdelta}) has been arranged to highlight that there is a linear relation between the model image intensities, $I_{l_k m_k}$, and the observed visibilities, $V_j\left( u_j,v_j,w_j \right)$. Denoting vectors and matrices with bold uppercase type, this may be written simply as
\begin{equation}
{\bf V} = {\bf M}{\bf I} \label{eq:sme}
\end{equation}
where
\begin{equation}
 {\bf M} \equiv M_{jk}= \frac{A_{l_k m_k}^{(j)}}{\sqrt{1-l_k^2-m_k^2}} e^{-2 \pi i  \left( u_j l_k+ v_j  m_k+w_j \left(\sqrt{1-l_k^2-m_k^2}-1\right).\right)} 
\label{eq:deltam}
\end{equation}
Generally the dimension of ${\bf I}$ is larger than that of ${\bf V}$, that is, the number of pixels is larger than the number of observed visibilities, so equation (\ref{eq:sme}) is under-constrained. To constrain the problem an additional constraint must be added to the system based on a priori knowledge. In this case it is assumed that the solution will be sparse, that is, have many zero components, and this assumption will be expressed by requiring that the solution have a minimal $L_1$ norm while still agreeing with the observed visibilities. To do this, a regularised error function is introduced of the form
\begin{equation}
E = |{\bf V} - {\bf M}{\bf I}|^2+\lambda \sum_k |I_k| \label{eq:cme}
\end{equation}
the deconvolution task is to search for the  ${\bf I}$ that minimises this error function. 

Note that while equation (\ref{eq:cme}) was formulated in terms of point sources, it can also be evaluated for any pixel shape for which an analytic Fourier transform of the pixel shape multiplied by a quadratic phase function may be found. We now derive the form for ${\bf M}$ for Gaussian shaped pixels for narrow field and wide field application.

\subsection{Gaussian pixels in the paraxial approximation} 
To model Gaussian shaped sources, a new class of pixel shapes is defined by
\begin{equation}
f_k(l,m)=  \frac{I_{l_k m_k}}{\sigma_k^2} e^{- \pi  {\left((l-l_k)^2+(m-m_k)^2\right)}/{\sigma_k^2}} \label{eq:gaus}
\end{equation}
which have been normalised so that the integral under the gaussian is one. We substitute equation (\ref{eq:gaus}) into equation (\ref{eq:samlm}), and a pre- and post-multiply by quadratic phase terms, leading to
\begin{equation}
\begin{array}{l}
V_j =  \sum_{k}\int{ \frac{A(l,m;Z_j,\chi_j)f_k(l,m)}{\sqrt{1-l^2-m^2}}}\times \\
\hspace{0.9cm}e^{  -2 \pi i  w_j \left( \sqrt{1-l^2-m^2}-1-\left(l^2+m^2\right)/2\right)}  e^{ i \pi w_j \left(l^2+m^2\right)}e^{-2 \pi i  \left(u_j l+v_j m\right)} dl\,dm.  
\end{array}
 \label{eq:samlmexp}
\end{equation}
We then Taylor expand the first phase term around the phase centre, leading to 
\begin{equation}
\begin{array}{l}
e^{  -2 \pi i  w_j \left( \sqrt{1-l^2-m^2}-1-\left(l^2+m^2\right)/2\right)} \\
\hspace{0.5cm}\approx 1+\frac{1}{4} i \epsilon ^4 \left(\pi  l^4 w+2 \pi  l^2 m^2 w+\pi  m^4 w\right)+O
\left(\epsilon^5\right)
\end{array}
\label{eq:taylorexp}
\end{equation}
where $l$ and $m$ are assumed of size $\epsilon$. Thus, to second order in $l$ and $m$, 
\begin{equation}
V_j =  \sum_{k}\int{ \frac{A(l,m;Z_j,\chi_j)f_k(l,m)}{\sqrt{1-l^2-m^2}} e^{ i \pi w_j \left(l^2+m^2\right)}e^{-2 \pi i  \left(u_j l+v_j m\right)}} dl\,dm.   \label{eq:samlm2}
\end{equation}
This approximation is equivalent to the well known paraxial approximation in optics, and leads to a phase error in the integrand of (\ref{eq:samlm}). For a representative $w_j=1000$ this is a phase error of $10^{-3}$ approximately 3 degrees from the pointing centre. It is also well known in Fourier optics that the quadratic phase term in equation (\ref{eq:samlm2}) represents a defocus - thus the $w$-term relates to a defocus between the dishes spaced at different depths relative to the pointing direction.  
Inserting the definition for the gaussian pixels, equation (\ref{eq:gaus}), into equation (\ref{eq:samlm2}), and making a further assumption that the direction dependent gains and projection factor do not vary significantly over a single Gaussian, we write
\begin{equation}
\begin{array}{l}
V_j =  \sum_{k} \frac{A(l_k,m_k;Z_j,\chi_j)}{\sigma_k^2\sqrt{1-l_k^2-m_k^2}}I_{l_k m_k} \times \\ \hspace{1.0cm} \int{  e^{- \pi  {\left((l-l_k)^2+(m-m_k)^2\right)}/{\sigma_k^2}}  e^{ i \pi w_j \left(l^2+m^2\right)}e^{-2 \pi i  \left(u_j l+v_j m\right)}} dl\,dm.   \label{eq:samlm3}
\end{array}
\end{equation}
This integral may be performed analytically, leading to and expression for ${\bf M}$ in equation (\ref{eq:cme}) for Gaussian pixels given by
\begin{equation}
\begin{array}{l}
{\bf M} \equiv M_{jk} =   \frac{A(l_k,m_k;Z_j,\chi_j)}{\sigma_k^2\sqrt{1-l_k^2-m_k^2}}
\frac{1}{1-i w_j \sigma_k^2} \times\\ 
\hspace{2.5cm}
e^{-2\pi i\frac{l_k u_j+m_k v_j}{1-i w_j \sigma_k^2}}
e^{i \pi w_j\frac{l_k^2 +m_k^2}{1-i w_j \sigma_k^2}}
e^{-\pi \sigma_k^2\frac{u_j^2 +v_j^2}{1-i w_j \sigma_k^2}}.
\end{array}
   \label{eq:samlm4}
\end{equation}
The three exponential terms in equation (\ref{eq:samlm4}) may be understood as follows. The first term is a modified linear phase term that corresponds to the spatial offset from the phase centre of the $k$-th gaussian pixel. The second term is a modified quadratic phase term, corresponding to the defocus due to the $w$ value of the $k$-th gaussian pixel. The final term is a modified gaussian, with scale $1/\sigma_k$ corresponding to the Fourier transform of the $k$-th gaussian pixel. In all cases, there is a modification by the denominator of $1-i w_k\sigma_k^2$, which mixes the real and imaginary parts of each term according to the amount of defocus and the scale of the gaussian. Taking the limit of equation (\ref{eq:samlm4}) with $w_j\rightarrow 0$, this leads to the Fourier transform of a Gaussian, as predicted from equation (\ref{eq:ftvis}). Taking the limit as $\sigma_k\rightarrow 0$ for all $k$, leads to the paraxial approximation of equation (\ref{eq:samlm}), as is to be expected.

Equation (\ref{eq:samlm4}) allows the prediction of the contribution of a extended source of emission to the visibilities measured by any baseline, taking into account the non-coplanar baseline effect and direction dependent antenna gains. The assumptions made are that the source has a Gaussian profile, and that the source is not so extended that the gains and the coordinate projection term vary significantly over the source.

\subsection{Gaussian pixels in a wide field}
The approximation in equation (\ref{eq:taylorexp}) limits the field of view of the image that the algorithm can be applied to. To avoid this, the phase offset at the centre of the Gaussian pixel may be preserved in the Taylor series expansion. In this case we have that
\begin{equation}
\begin{array}{l}
{\bf M} \equiv M_{jk} =   \frac{A(l_k,m_k;Z_j,\chi_j)} {\sigma_k^2\sqrt{1-l_k^2-m_k^2}}e^{  -2 \pi i  w_j \left( \sqrt{1-l_k^2-m_k^2}-1-\left(l_k^2+m_k^2\right)/2\right)} 
\times\\ 
\hspace{2.0cm}
\frac{1}{1-i w_j \sigma_k^2} 
e^{-2\pi i\frac{l_k u_j+m_k v_j}{1-i w_j \sigma_k^2}}
e^{i \pi w_j\frac{l_k^2 +m_k^2}{1-i w_j \sigma_k^2}}
e^{-\pi \sigma_k^2\frac{u_j^2 +v_j^2}{1-i w_j \sigma_k^2}}
.
\end{array}
   \label{eq:samlm5}
\end{equation}

This form of ${\bf M}$ will be suitable for any field of view and is only limited by the Taylor expansion of the Gaussian pixel itself, i.e. as long as a single Gaussian pixel does not subtend an angle over which $w_j \sqrt{1-(l-l_k)^2-(m-m_k)^2}$ varies significantly. Equation (\ref{eq:samlm5}) requires more computation than equation (\ref{eq:samlm4}), however it is the form of the equation required for all-sky coordinate systems.

\section{Implementation of SL1M} \label{sec:fista}

The SL1M algorithm is an algorithm for image deconvolution which is represented as the solution of equation (\ref{eq:cme}) with ${\bf M}$  given by either equation (\ref{eq:deltam}) for delta function pixels or by equations (\ref{eq:samlm4}) or (\ref{eq:samlm5}) for Gaussian pixels.

Equation (\ref{eq:cme}) is equivalent to the minimisation problem treated in \cite{Li:2011p1778}. However in Li et al. the transformation between measured visibilities and source pixels is made through a Fourier transform of gridded visibilities which requires that the pixels be regularly spaced, and that the visibilities be transformed into the $w=0$ plane through a gridding operation. In contrast, for SL1M, the matrix {\bf M} represents an arbitrary mapping between source pixels and antenna gains and is explicitly evaluated for each visibility and pixel pair.

Because the minimisation problems are of the same form, the same numerical methods for solving the minimisation, namely the Fast Iterative Shrinkage-Thresholding Algorithm (FISTA) from \cite{Beck:2009p2984}, may be used.  Given the maximum eigenvalue of the matrix ${\bf M}$ the FISTA algorithm guarantees $1/k^2$ convergence, where $k$ is the number of iterations of the algorithm. This is unlike deconvolution algorithms based on CLEAN, where no such guarantee of convergence can be made. Furthermore, the parameter $\lambda$ in equation (\ref{eq:cme}) is the only major free parameter (excluding parameters related to the sampling pattern in the image space). This parameter controls the trade-off between errors in reconstructing the observed visibilities, and enforcing the sparsity of the reconstructed solution, and may be set based on the expected brightness of the sources and the noise in the sampled visibilities. Note that other algorithms exist for performing this minimisation, some of which show faster convergence for a variety of applications (\cite{Becker:2011p3142}). As the FISTA algorithm is considered a gold standard for L1 minimisation problems, and because it has previously been shown to work in the radio synthesis context (\cite{Li:2011p1778}), we adopt it here, though other minimisation approaches may be faster. For a detailed examination of many algorithms related to L1 minimisation, the reader is referred to \cite{Bach:2011p3079}. 

The FISTA algorithm which produces a sequence of estimates, ${\bf I}_k$, is shown here:

\begin{tabular}{ll}
{\bf Input:} & $L$ - maximum eigenvalue of ${\bf M}^*{\bf M}$ \\
                 & $\lambda$ - regularisation parameter \\
		 & ${\bf V}$ - values to be fit \\
{\bf Step 0:} & ${\bf y}_1 = {\bf I}_0 = 0$, \\
		    & $t_1 = 1$ \\
{\bf Step k:} & ${\bf I}_k = T^{\lambda/L}\left( {\bf y_k} - \frac{1}{L} {\bf M}^*({\bf M} {\bf y}_k - {\bf V})\right)$	 \\
	& $t_{k+1} = \frac{1+\sqrt{1+4 t_k^2}}{2}$ \\
	& ${\bf y}_{k+1} = {\bf I}_k + \frac{t_k-1}{t_{k+1}}\left({\bf I}_k - {\bf I}_{k-1}\right)$\\
\end{tabular}

This algorithm can be terminated when an iteration leads to a sufficiently small change in the total error $E$, given by equation (\ref{eq:cme}), or when there is a sufficiently small change in the number of non-zero entries in ${\bf I}_k$.  The positivity enforcing thresholding operation is defined by
\begin{equation}
T^d({\bf x}) = \left\{ \begin{array}{lr}x_i - d & x_i>d\\0 & x_i \leq d \end{array}. \right.
\label{eq:threshp}
\end{equation}

There are three key parts to the algorithm. The first key part of the algorithm is the step which performs the L2 minimisation. This is a gradient descent step, where the derivative of $\left|{\bf M} {\bf y}_k - {\bf v}\right|^2$ with respect to ${\bf y}_k$ is evaluated. This derivative is ${\bf M}^*({\bf M} {\bf y}_k - {\bf v})$, which essentially back projects the residual visibility errors into the image domain. This term is scaled by $1/L$, where $L$ is the maximum eigenvalue of ${\bf M}^*{\bf M}$, which ensures that the step is small enough not to diverge from the correct solution.  Hence $L$ determines how quickly the algorithm converges.

The second key part to the algorithm is the threshold-shrinkage step. This moves the solution closer to the L1 minimised solution by removing small (and negative) values from the solution. The third key part of the algorithm is the update step, where a particular linear combination of the previous steps is used to guarantee convergence at a rate of $1/k^2$.

Finally, it is also important to note that the FISTA algorithm is not monotonic. That is, an iteration may lead to an increase in the value of the error term, and this is not indicative of the convergence of the algorithm. A monotonic version of FISTA, MFISTA, is presented in \cite{Beck:2009p3004}, but it is not used here as it requires an additional evaluation of the matrix ${\bf M}$.

\subsection{On-the-fly computation versus in-memory computation}

Direct application of the FISTA algorithm to equation (\ref{eq:cme}) requires requires repeated evaluation of the matrix ${\bf M}$ and its transpose. For the case where the pixels are on a rectangular grid, the matrix ${\bf M}$ is of dimension $N_v \times N_l N_m$, which, for a reasonable size observation can be $500,000 \times 1,000,000$. If this matrix were to be stored in memory using a 4-byte floating point number to represent the real and imaginary elements it would require over 4 terabytes of RAM.  Thus, while this matrix has a simple form, using it in an iterative algorithm represents a very large numerical problem. 

However, there is an alternative approach to storing all this data. In this approach the elements of the matrix are recalculated as they are needed and are not stored. This technique turns the solution of equation (\ref{eq:cme}) from a large memory, high memory-bandwidth task into a low-memory, processor intensive computational task that is extremely well suited to modern multi-core hardware due to the highly parallelisable nature of the problem. 

We have implemented the SL1M algorithm through this method of explicit evaluation of the components of ${\bf M}$ from their analytical representation (given by equations (\ref{eq:deltam}), (\ref{eq:samlm4}) or (\ref{eq:samlm5})). This involved implementation of the FISTA algorithm using C++ and CUDA on GPGPU hardware, along with an algorithm to calculate the largest eigenvalue of ${\bf M}^*{\bf M}$. The specific hardware used to run this code were 2 Fermi class M2050 GPUs attached to a cluster processor available on Amazon Web Services Elastic Cloud Compute platform. The GPUs have a maximum floating point performance of 500GFLOPs each. An evaluation of a single term of ${\bf M}$ takes approximately 30 FLOPS using single precision floating point arithmetic and fast sincos and sqrt primitives. This means theoretical peak performance in evaluation of entries of ${\bf M}$ is around 33 billion entries per second. Real world performance is 99 per cent of the theoretical maximum due to the independence of the entries, and the low amount of memory bandwidth required for the calculation. For this architecture, the calculation is distributed across approximately 21,000 threads on each GPU. For the case of a matrix of size $500,000 \times 1,000,000$, evaluation of a single FISTA step takes 30 seconds. An image deconvolution may take hundreds of steps of the FISTA algorithm to converge, leading to run times in the order of hours, depending on the nature of the problem.

\begin{figure} 
%\centering
\includegraphics[width=84mm]{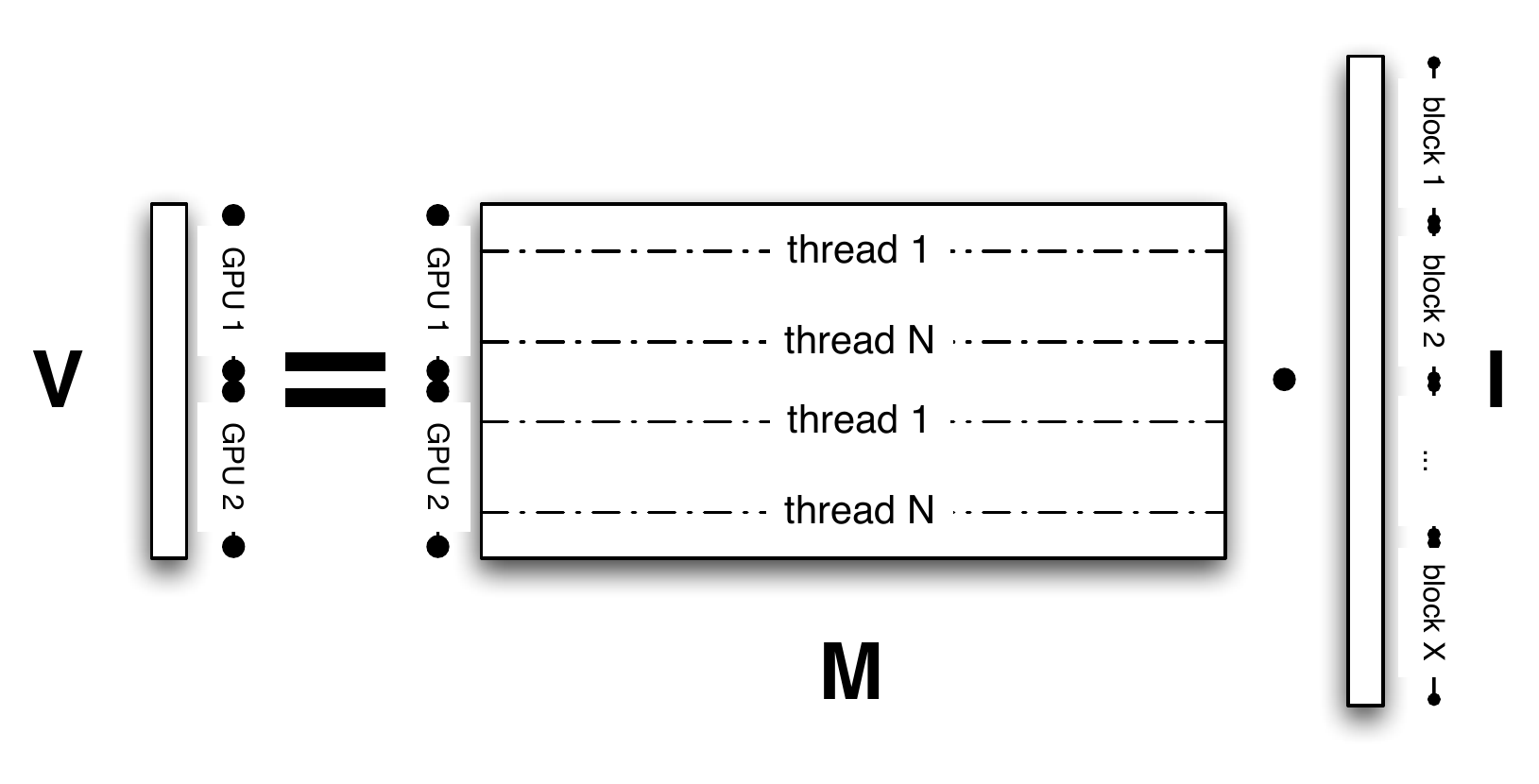}
\caption{Parallelisation scheme for a matrix vector multiply in SL1M to two GPUs on a single host. Each entry of M is calculated as required to match the pixel position and the visibility being calculated. }
\label{fig:parallel}
\end{figure}

The parallelisation scheme used for two GPUs on a single host is shown diagrammatically in Fig. \ref{fig:parallel} for a multiplication between the image vector and the matrix ${\bf M}$. Half the calculation is done on each GPU, and the image vector is divided into blocks to aid efficient memory access.  Further parallelisation of the algorithm is possible by distributing to multiple machines. This may be achieved by extending the scheme of Fig. \ref{fig:parallel} where the entire image to be updated is shared between machines via the network, or by splitting the image pixels between hosts. In this case, each host on the network has a subset of the image that it calculates with, and for each matrix multiply, it only calculates the elements of the matrix corresponding to the pixels it contains and then distributes the results to the host node over the network. Scaling to a cluster of GPU machines is feasible with this technique and would reduce the computing time per iteration roughly linearly in the number of machines, with some overhead for the network communication. This approach has not yet been implemented. The code which calculates the results shown here is freely available online\footnote{http://github/StephenJHardy/SL1M}.

As this is a new algorithm, an emphasis has been placed on demonstrating the accuracy and reliability of the deconvolution result, and not on the performance of the code. As such, all results reported here are run over many hundreds, and sometimes thousands, of steps of the FISTA algorithm. This is not always required, particular for real noisy data as shown in Sections \ref{sec:ngc5921} and \ref{sec:ngc2403}. Further work in algorithmic optimisation is discussed in Section \ref{sec:accel}. 

The approach used here of calculating the explicit transformation between the image pixels and the observed visibilities as they are required for the calculation, rather than pre-calculating and storing them in RAM, could be applied to other deconvolution algorithms. The requirements are that there is an analytic form for the transformation, and that the algorithm require only evaluations of ${\bf M}$ or its transpose. Methods that require the solution of sets of linear equations as part of their optimisation algorithms cannot make use of these techniques.

\section{Results}

\subsection{Synthetic data}
To begin the evaluation of the performance of SL1M, we initially test it on synthetic data, both with and without noise. This is followed by the analysis of two real data sets drawn from observations of NGC5921 and NGC2403 by the VLA telescope. 

\subsubsection{Point sources} \label{sec:ps}
The initial synthetic dataset consists of data generated by simulating 50 point sources randomly distributed over a 7 degree field of view which is represented by a 1024 $\times$ 1024 image with 30" pixel spacing. The sources are distributed over the inner 80 per cent of the image, and have strengths ranging from 0.2 to 2.0 in arbitrary units. Visibilities are generated by simulating the dish distribution of the full ASKAP telescope (\cite{Deboer:2009p3117}), with 36 dishes, but for only the central beam, a single polarisation and a single channel at the HI wavelength. The primary beam of the telescope is also not modelled. Visibilities are calculated for a one hour period, sampling every minute, leading to a total of 37,800 visibility records. The centre of the field is assumed to be at a declination $-22.5^o$ and at zenith at the start of the observation.

Initially, we test the effectiveness of the algorithm in the absence of measurement noise. To do this we run SL1M on the simulated visibilities for a variety of values of $\lambda$ until the total error changes no more than 1 part in $10^6$ or until 7000 iterations were reached. The results of these tests are shown in Table \ref{tab:pointsourcenonoise}. Note that for the $\lambda = 1.0$ test, the reconstruction error for the 50 non-zero sources was less that $4 \times 10^{-4}$. This demonstrates similar accuracy to previous applications of compressive sensing (e.g. \cite{Candes:2005p3330} ). It is worth noting that while the number of observed visibilities (37,800) is larger than the number of non-zero samples being reconstructed (50), the number of pixels in the solution is larger still (1,048,576). 

\begin{table}[!th]
\centering
\begin{tabular}{|c|c|c|c|c|c|}
\hline
$\lambda$   & Iter & $L_2$ term                 & $L_1$ term    &  error  & error \\
& & & & (RMS) & (max)\\
\hline
$10^{-2}$             & 7000        & $8.3 \times 10^{-3}$    & 1448         &  $8.1 \times 10^{-3}$    & 1.61\\
$10^{-1}$              & 7000	      & $4.7 \times 10^{-2}$    & 655.5           &   $4.6 \times 10^{-3}$     & 0.90 \\
1              & 4992	      & $8.6 \times 10^{-1}$    & 61.58             &   $5.5 \times 10^{-6}$   & $4.3 \times 10^{-4}$ \\
$10^1$                & 1860	      & 4.8                             & 60.62             &   $3.9 \times 10^{-6}$   & $1.1 \times 10^{-3}$ \\
$10^2$         & 1367	      & 9.8                             & 60.60             &   $1.2 \times 10^{-5}$   & $4.2 \times 10^{-3}$ \\
$10^3$           & 454	      & 460                           & 59.90             &   $1.1 \times 10^{-4}$     & $2.8 \times 10^{-2}$ \\
\hline
\end{tabular}
\caption{Deconvolution results for the SL1M algorithm for a variety of regularisation parameters ($\lambda$) run against 37,800 vsibilities simulated from a test image of 50 point sources on a 1024x1024 pixel grid. No noise has been added to the visibilities. The runs were limited to 7000 iterations or a change in total error of 1 part in $10^6$. The error terms for equation (\ref{eq:cme}) are shown separately in the third and fourth columns. The reconstruction errors are based on comparison with the original image. Note that the true $L_1$ norm of the input data was 60.5996.}
\label{tab:pointsourcenonoise}
\end{table}

\begin{figure}
\centering
\includegraphics[width=84mm]{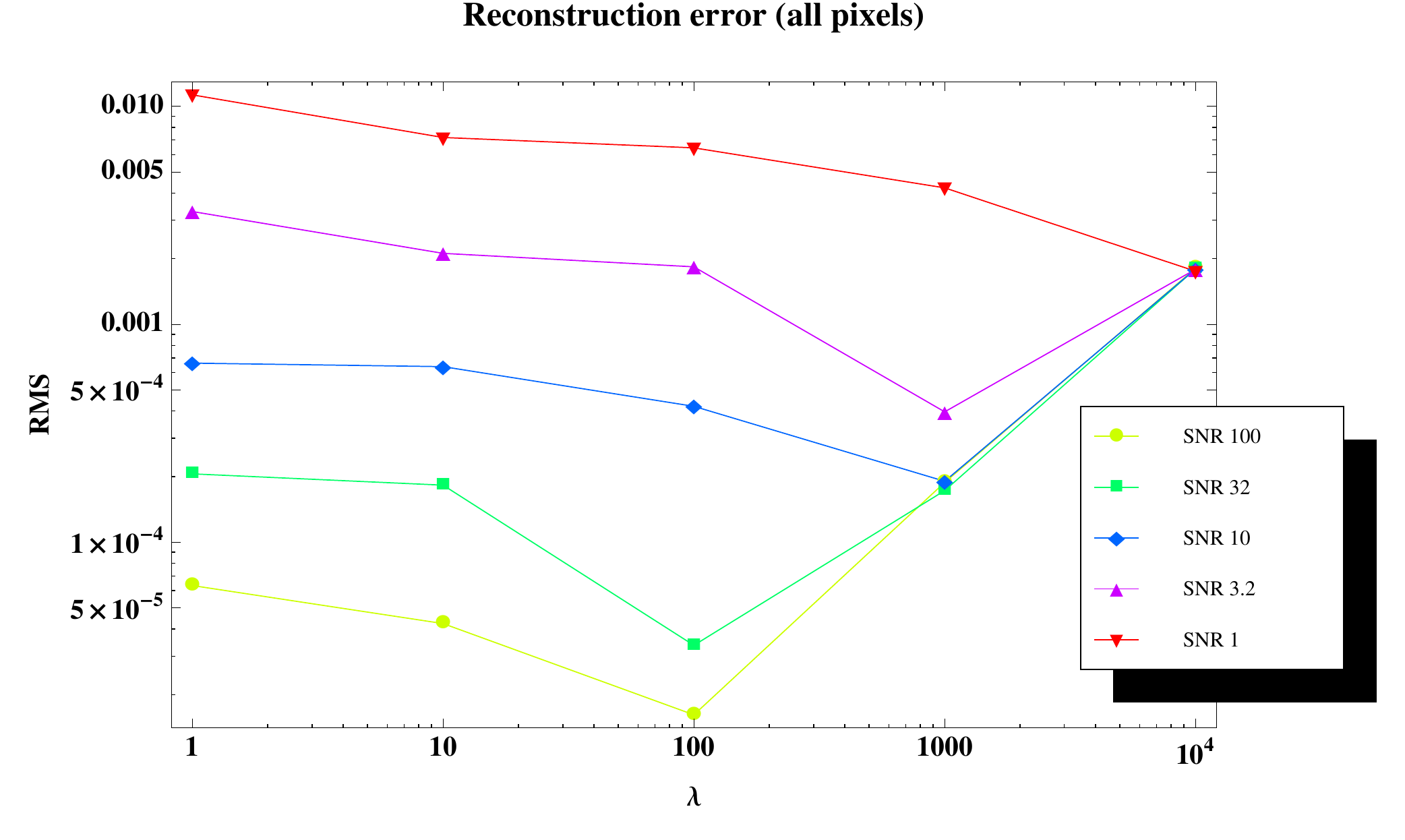}
\includegraphics[width=84mm]{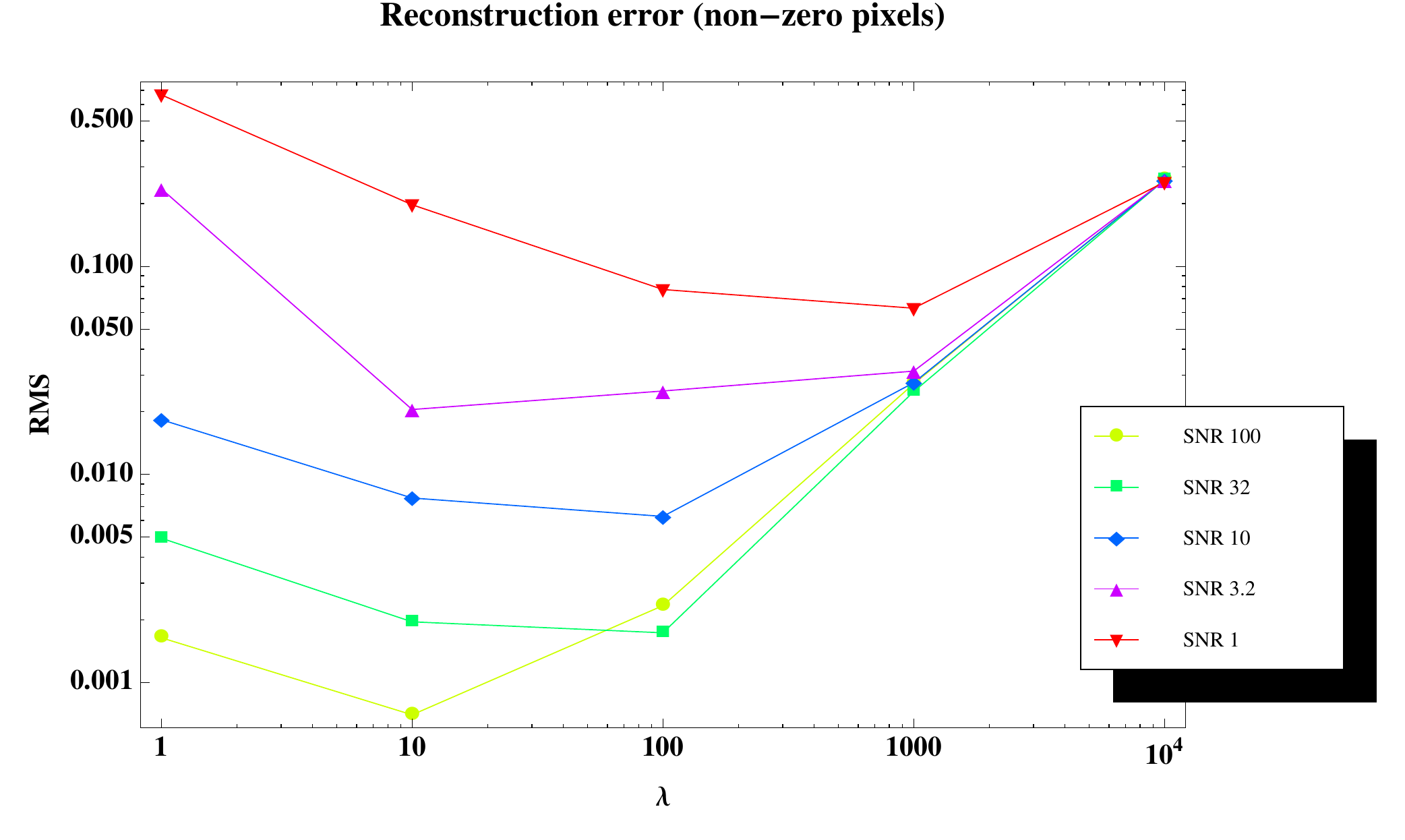}
\caption{RMS error in reconstructed point source image  as a function of regularisation parameter $\lambda$, for 5 different noise levels. The first plot shows the RMS error level for all pixels. The second plot shows the RMS noise level for just the non-zero pixels. }
\label{fig:pointreconerror}
\end{figure}

Next the effectiveness of the algorithm in the presence of noise is tested on the same dataset, but with additive noise combined with the visibility data. We added gaussian noise with zero mean and a specified standard deviation to both the real and imaginary parts of all the visibilities, and the reconstruction algorithm was run for 5 different values of $\lambda$. The standard deviations were specified so that the signal to noise ratios were 100, 31.6, 10, 3.16 and 1. To evaluate the performance of the deconvolution algorithm, the RMS difference between the reconstructed image and the original image is plot as a function of $\lambda$, for the different noise levels, in Fig. \ref{fig:pointreconerror}. This figure shows that good reconstruction results are possible to a signal to noise ratio of at least 1. Even for this case the RMS error is 0.017, which is much smaller than the non-zero pixels which have amplitudes of between 0.2 and 2.0. It is also important to note that, as the noise becomes progressively worse, the best reconstruction is obtained with a higher value of $\lambda$. This is because the $L_2$ term in equation (\ref{eq:cme}) increases relative to the $L_1$ term as the noise increases, so $\lambda$ must be increased to avoid fitting the noise. The second panel of Fig. \ref{fig:pointreconerror} shows the RMS error for the non-zero pixels. This may be a better figure of merit than the total RMS error as these are the pixels that have physical significance in real data. In this case, the RMS is lower for lower values of $\lambda$ than in the first panel. This may be explained, as the $L_1$ term of equation (\ref{eq:cme}) penalises higher values of the solution. Thus, there is a trade off between suppressing noise in background regions, and maintaining the accuracy of the solution in regions where there is signal.

\subsubsection{Extended emission}

\begin{figure}
\centering
\includegraphics[width=84mm]{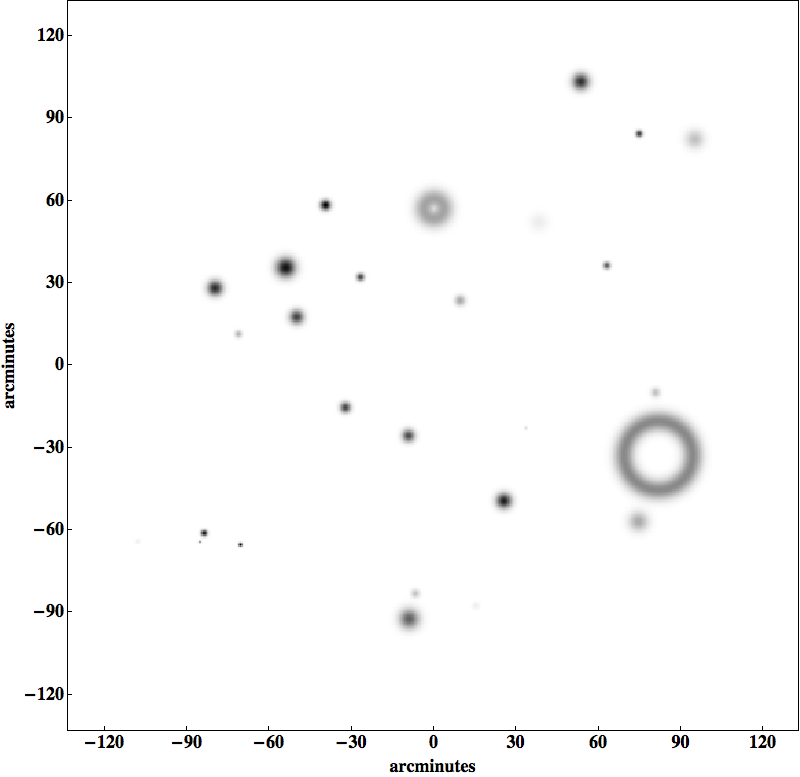}
\caption{Synthetic test image with 25 gaussian sources and two ring structures on a 512 $\times$ 512 pixel grid with 30 arc second pixel spacing. There are 16098 pixels with intensity more that 0.001 times the maximum intensity.}
\label{fig:synthim}
\end{figure}

The previous test cases were ideal for the algorithm under investigation - the source image consisted of delta-functions, which matched the emission model. In this section, a deconvolution task using a synthetic image with extended emission is investigated. The image used is shown in Fig. \ref{fig:synthim}. It consists of a number of gaussian shaped sources, supplemented with two rings. Visibilities for the ASKAP telescope are simulated under the same conditions as section \ref{sec:ps}, and noise is added to the visibilities giving a signal to noise ratio of 1. The performance of the algorithm in reconstructing the image is investigated in for six different data lengths, ranging from 25000 to 150000 visibilities. 

\begin{figure}
\centering
\includegraphics[width=84mm]{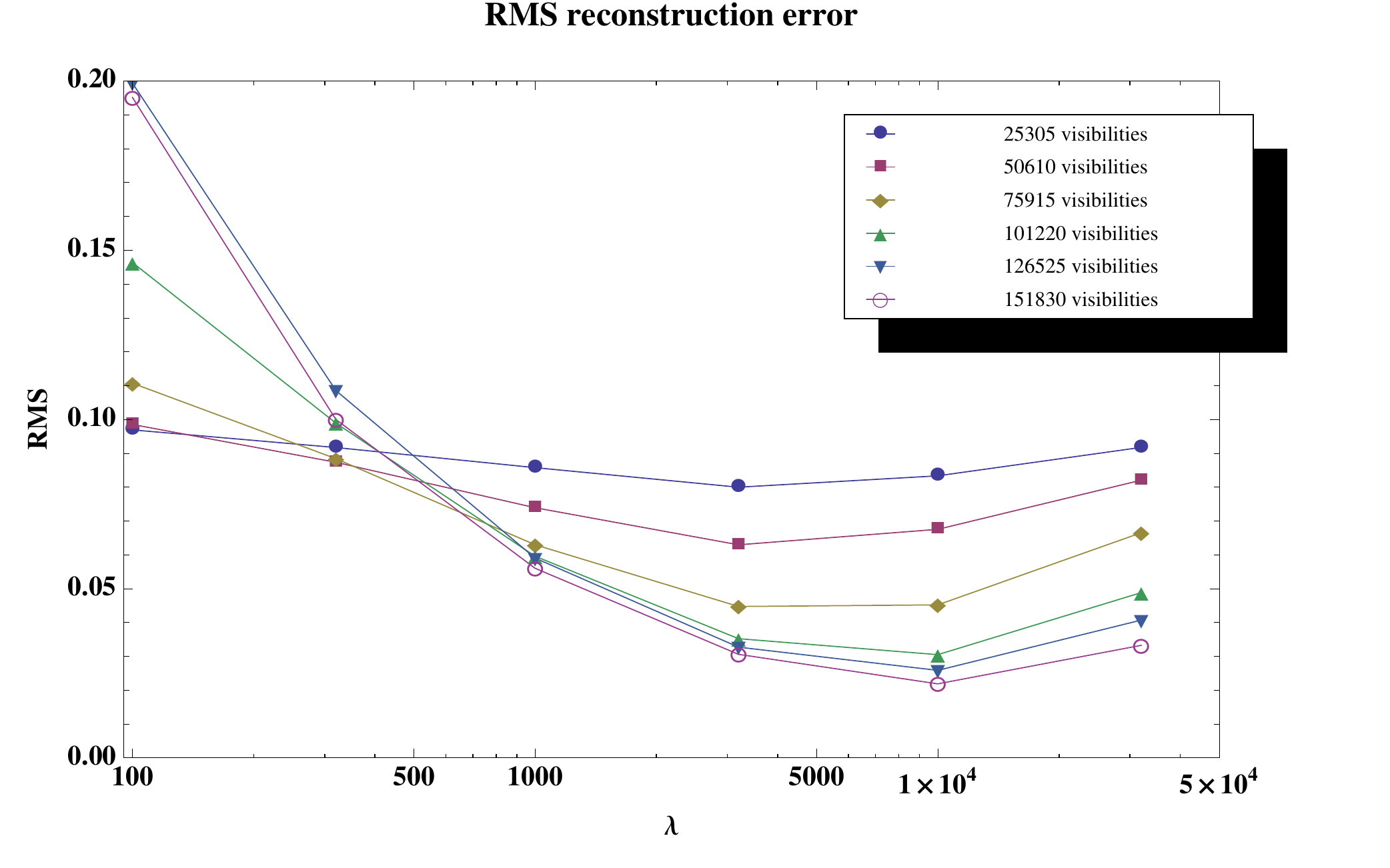}
\caption{RMS reconstruction error of a synthetic test image with 25 gaussian sources and two ring structures on a 512 $\times$ 512 pixel grid with 30 arc second pixel spacing.  Intensities of the source image vary from 0 to 2 in arbitrary units. }
\label{fig:rmsvis}
\end{figure}

The results of the RMS accuracy of the reconstruction are shown in Fig. \ref{fig:rmsvis} as a function of $\lambda$ for each of the data lengths. Note that the minimum error occurs at increasing values of $\lambda$ as the data length increases. Similarly to the case for point sources, this is due to the increase in the $L_2$ error as more data is introduced relative to the fixed $L_1$ norm of the solution. 

\begin{figure} 
\centering
\includegraphics[width=84mm]{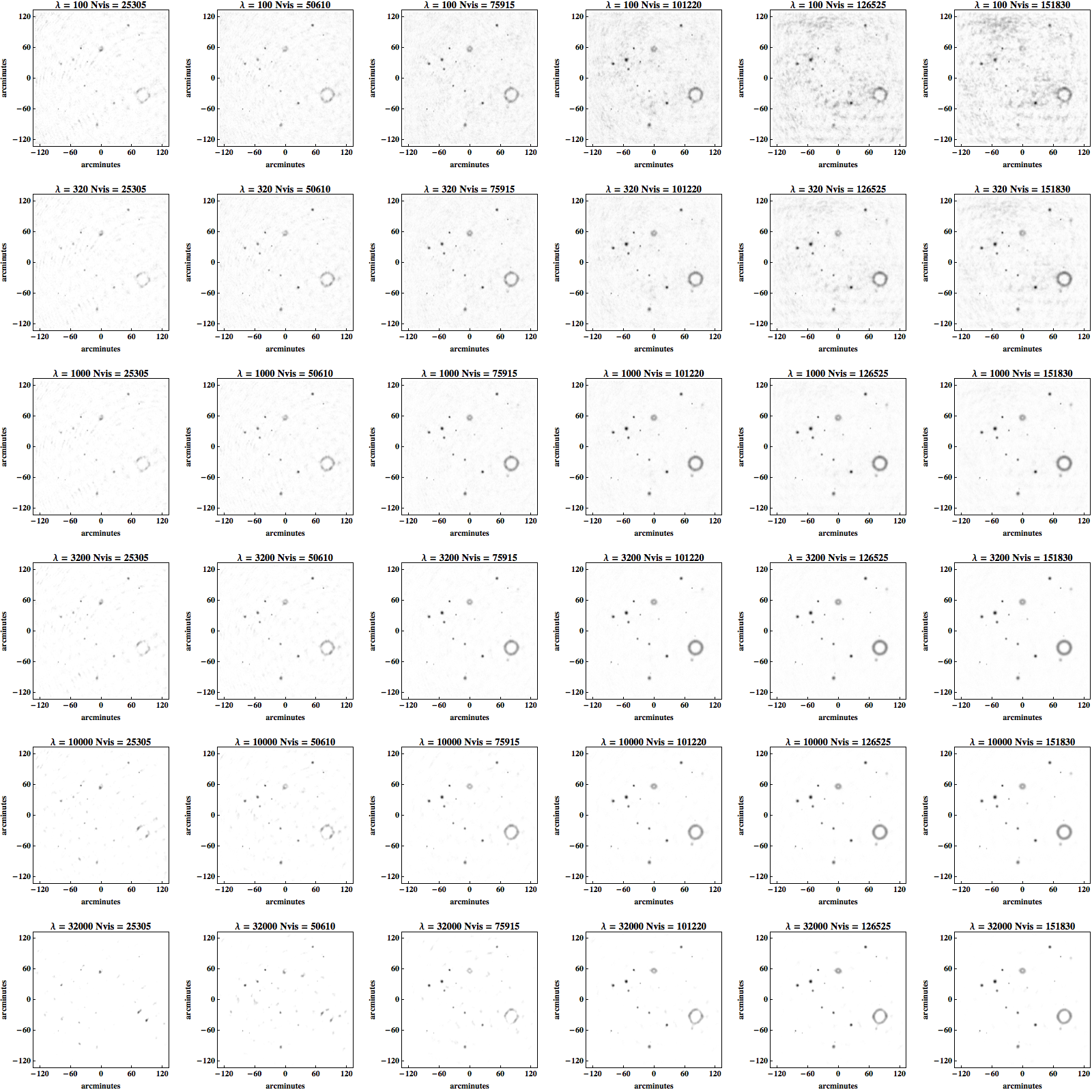}
\caption{Reconstructed images from visibilities calculated from the test image in Fig. \ref{fig:synthim}. From left to right, the number of visibilities increases from 25,305 to 151,830. From top to bottom, $\lambda$ takes the values 100, 320, 1000, 3200, 10000, and 32000. Note that these images are the final result of the FISTA algorithm - they have not been convolved with the synthesised beam of the telescope, and no residuals have been added.}
\label{fig:reconims}
\end{figure}

The actual reconstructed images are shown in Fig. \ref{fig:reconims}. This figure shows that the algorithm over-smooths the data for higher values of $\lambda$ and low numbers of visibilities (bottom left of the grid), and it over fits the noise for lower values of $\lambda$ and higher numbers of visibilities. Reconstructions of increasingly better quality occur for larger datasets, corresponding to the minima of Fig. \ref{fig:rmsvis}.
\newpage

\subsection{NGC5921} \label{sec:ngc5921}
To test with real data, we deconvolve the NGC5921 dataset that is distributed as a tutorial with the CASA radio astronomy software package\footnote{http://casa.nrao.edu}. This dataset consists of 63 channels of LL and RR polarisations, taken with the 27 telescopes of the VLA in a band centred on HI with a total bandwidth of 1.6MHz. In total, 11,934 visibilities for each channel were measured. The visibilities were calibrated and continuum subtracted according to the recommendations in the CASA software tutorial and exported for analysis. Only unpolarised emission was considered, so the LL and RR polarisation data were added to produce the visibilities input to SL1M.

\begin{figure} 
\centering
\includegraphics[width=28mm]{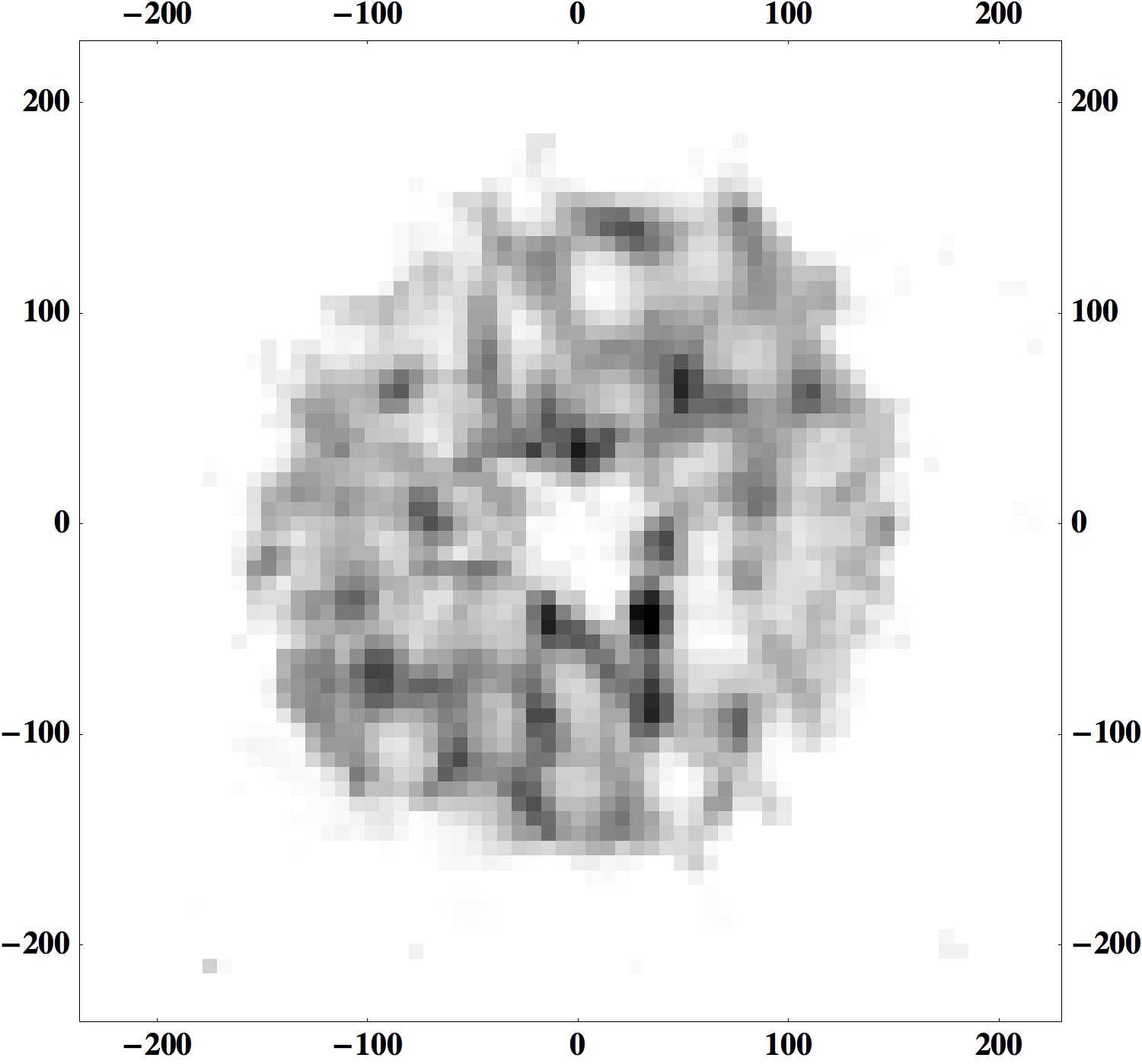}
\includegraphics[width=28mm]{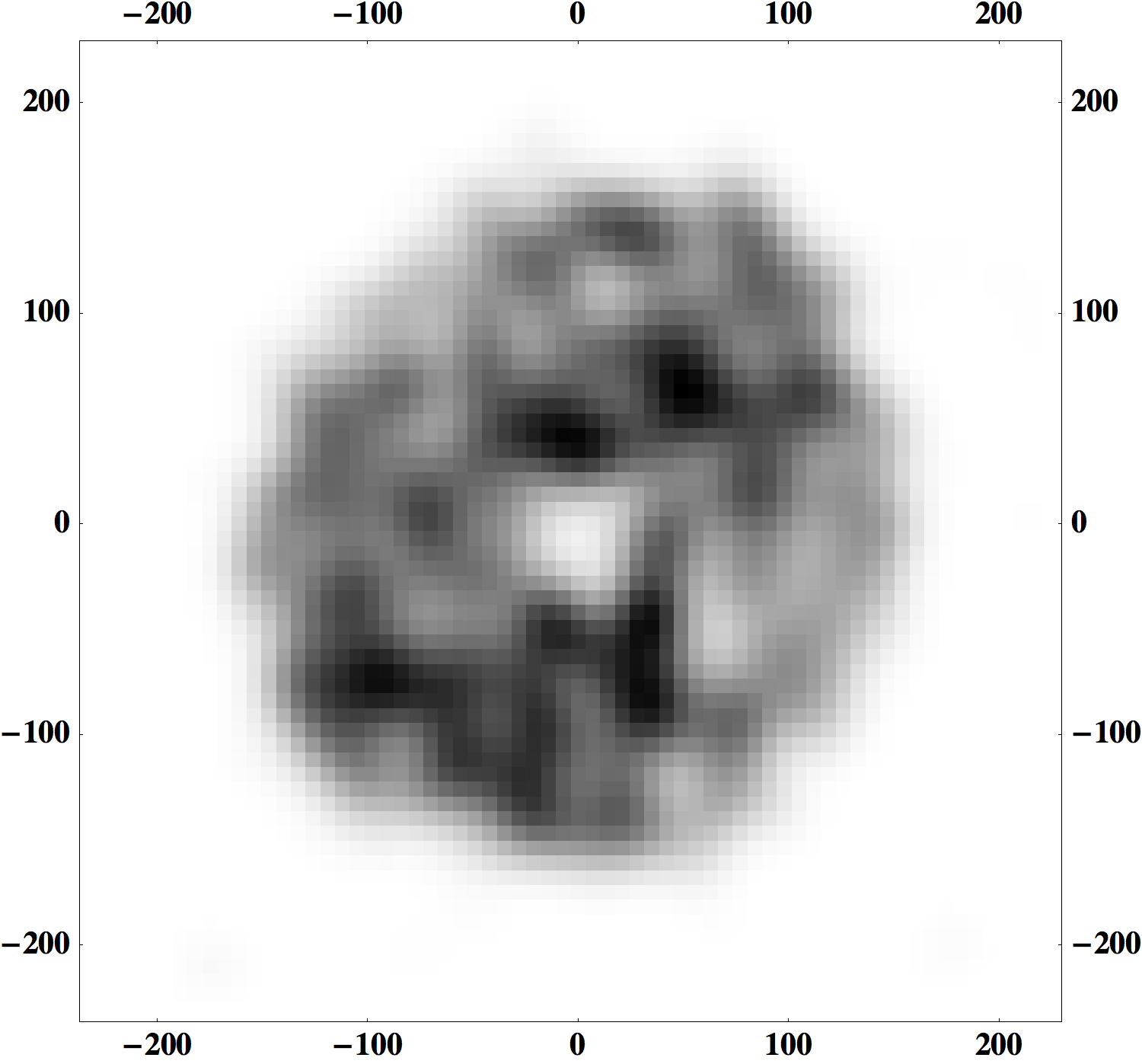}
\includegraphics[width=26mm]{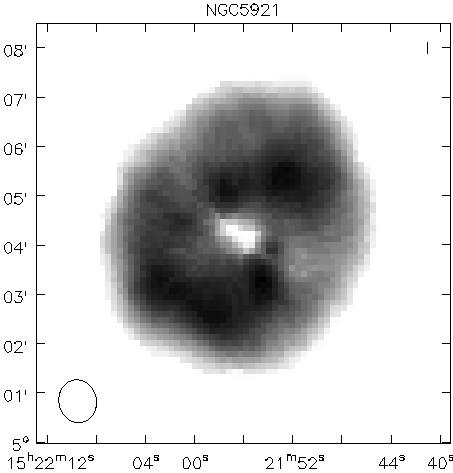}
\caption{Deconvolved image of NGC5921 reconstructed at 7 arc seconds per pixel. The first panel shows inner 64x64 portion of the sum of channels 10 to 50 of the 256x256 raw output from the SL1M algorithm. The second panel shows the same region as the first, but convolved with a 28 arc second gaussian to approximate the synthesised beam of the telescope. The third panel shows a CLEAN based reconstruction from the CASA software package. }
\label{fig:reconngc5921}
\end{figure}

The result of applying the SL1M algorithm over all 63 channels of the data with $\lambda=120$ are shown in Fig. \ref{fig:reconngc5921}.  The image is deconvolved to a 256 $\times$ 256 grid with a pixel spacing of 7 arc seconds. Each channel was processed until the relative change in the total errors was less than $10^{-9}$. Generally only around 200 iterations per channel were necessary, and this took around 30 seconds per channel. The first image shows the sum of the direct output of the SL1M algorithm for channels 10-50, and the second panel shows the same convolved with a gaussian approximation to the synthesised beam of the telescope. The third panel shows the corresponding CLEAN image generated using the CASA software using the default configuration, at the same pixel spacing as the SL1M algorithm. 

\begin{figure} 
\centering
\includegraphics[width=67mm]{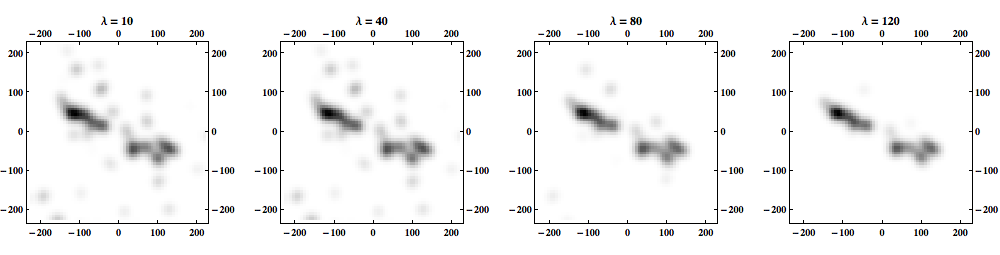}
\includegraphics[width=17mm]{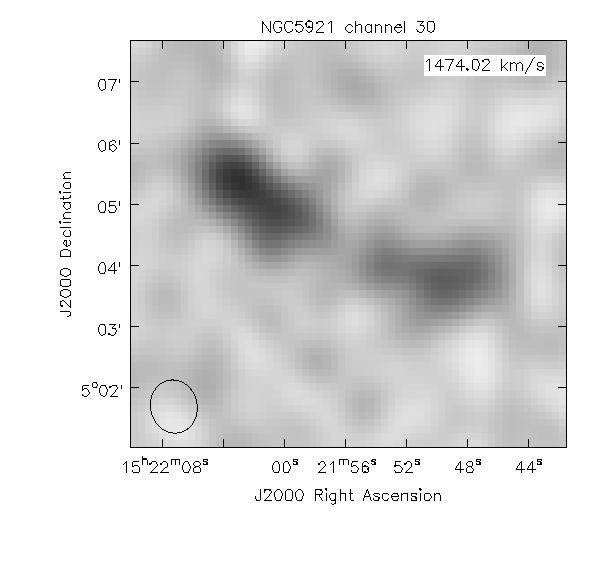}
\caption{Inner 64$\times$64 pixel area of a single channel (channel 30) of the deconvolved image of NGC592 reconstructed with 4 different values of $\lambda$ - 10, 40, 80, and 120.  The lower panel shows the result of CLEAN for this area. Note that the CLEAN image has had the residuals after clean processing added back into the image. This is currently not possible with the SL1M algorithm. }
\label{fig:channel30}
\end{figure}

A single channel of the result of the SL1M algorithm for 4 different values of $\lambda$ is shown in Fig. \ref{fig:channel30}. Increasing the value of $\lambda$ increases the strength of the $L_1$ minimisation term, thereby decreasing the noise in the reconstructed image. 

\subsection{NGC2403} \label{sec:ngc2403}
As a larger test, we deconvolve the NGC2403 dataset that is also distributed with a tutorial\footnote{http://casa.nrao.edu/Doc/Scripts/ngc2403\_tutorial.py} for the CASA software. This dataset has 432,783 visibility records for 127 channels starting at 1418.25MHz with a channel bandwidth of 24.414kHz taken by the VLA. The synthesised beam size is around 12 arc seconds and the object is around 35 arc minutes across. For this test, the image is deconvolve onto a 1024 $\times$ 1024 pixel grid with a pixel size of 2 arc seconds. Again, the data included LL and RR polarisation measurements which were summed before deconvolution. This dataset includes some records affected by interference, and the noisy records were flagged and removed before deconvolution. As per the CASA tutorial, calibration and continuum subtraction were performed, with channels 21-30 and 92-111 used for continuum estimation. The execution time for a single channel for this dataset is 1.5 hours, and the reduction time for the 61 line channels was around 90 hours.

\begin{figure}
\centering
\includegraphics[width=84mm]{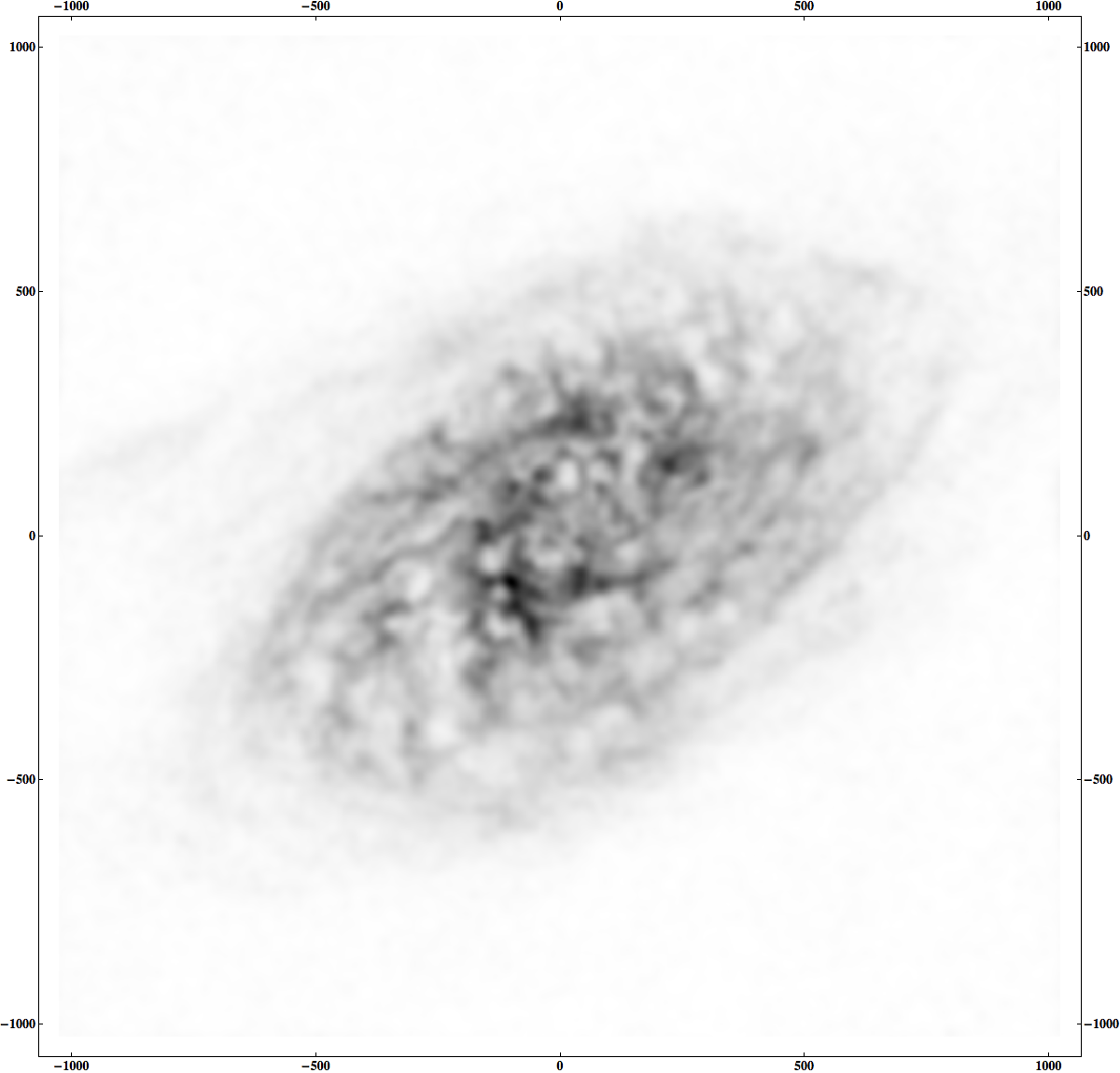}
\caption{Deconvolution result for NGC2403. This image is the sum of channels 31 to 91 and is the output of the SL1M algorithm with $\lambda=660$ convolved with a 12 arc second Gaussian. }
\label{fig:2403sum}
\end{figure}

The image generated from combining the deconvolved images from channels 31 to 91, and convolving with a 12 arc second Gaussian are shown in Fig. \ref{fig:2403sum}.

\subsection{Analysis at different scales}

\begin{figure}
\centering
\includegraphics[width=84mm]{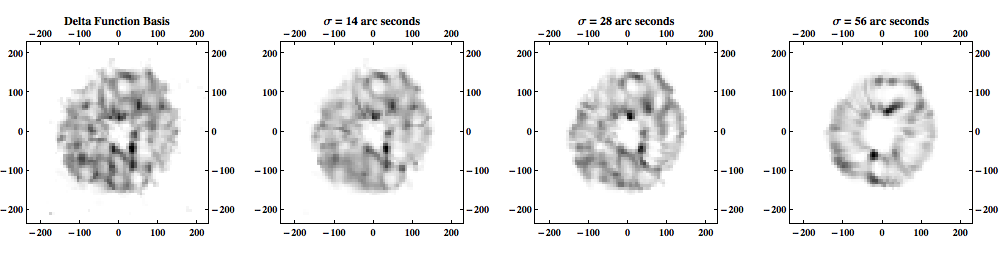}
\caption{Output from the SL1M algorithm applied to the NGC5921 dataset with different sized pixels. The first panel uses delta function pixels; the second panel uses Gaussian pixels 14 arc seconds (2 pixels) across; the third panel, 26 arc seconds; and the 4th panel, 56 arc seconds. Note that these images have not been convolved with the Gaussians corresponding to the pixel size, or with the synthetic beam of the telescope. }
\label{fig:multiplescales}
\end{figure}

To demonstrate processing at different scales, the NGC5921 dataset analysed in Section \ref{sec:ngc5921} is deconvolved with gaussian basis functions of different sizes based on equation (\ref{eq:samlm4}). The results of this analysis are shown in Fig. \ref{fig:multiplescales}. Larger scale pixels show correspondingly less detail, as might be expected. Also, the largest scale clearly shows the emission in each channel is perpendicular to the rotation of the galaxy. 

Multi-scale methods generally operate on multiple scales simultaneously, and this could be achieved here by having pixels with different scales in the same SL1M run. It is also possible to approximate the Isotropic Undecimated Discrete Wavelet transform used by \cite{Li:2011p1778} as the difference of two Gaussian kernels. 

Here, the ability to process with different scales allows is used to demonstrate an acceleration strategy that can greatly reduce the overall processing time for the method.

\subsection{Acceleration strategies} \label{sec:accel}

For the SL1M algorithm, the flexibility in pixel placement and scale and the direct nature of the solution method come at a significant computational cost. Current methods such as CLEAN use the FFT to transform between the visibility and image domains, which involves two steps - gridding, which takes time proportional to the number of visibilities, $N_{vis}$; and the FFT itself, which scales with the number of pixels, $N_p$, as $N_p \log N_p$. On the other hand. SL1M scales as $N_p N_{vis}$. As $N_{vis} >> \log N_p$, this takes significantly more computation. However, due to the flexibility of the approach, a number of other strategies can be taken to improve the computational complexity. 

The first method for reducing computational cost is to use the dirty image as the initial condition from the SL1M algorithm. This reduces the number of iterations required for each run to converge, though it does not improve the processing speed of each step.

A second method for reducing computational cost would be to work in a coarse-to-fine strategy. That is, to solve the equation on a coarse pixel grid, then to double the resolution and upscale the previously calculated solution. This method also reduces the number of iterations required to reach convergence, though does not change the order of complexity of the solution, as the final stage still requires a calculation of all the pixel against all of the visibilities.

\begin{figure}
\centering
\includegraphics[width=84mm]{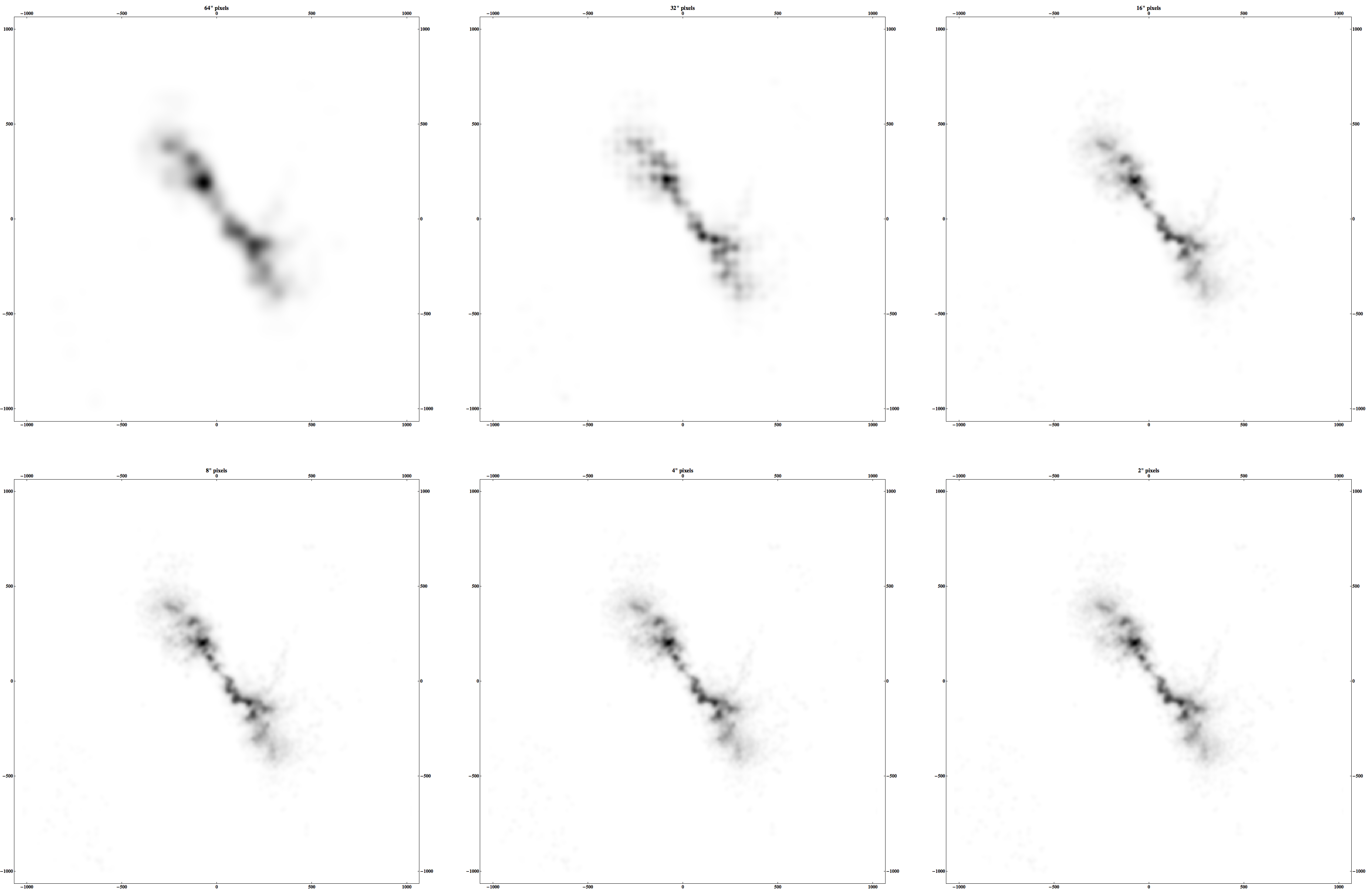}
\caption{Deconvolution result from a adaptive quadtree refinement process for channel 63 of the NGC2403 dataset. The first panel shows the result of the SL1M algorithm at the lowest resolution of 32x32 64" pixels. The second panel shows the result of the SL1M algorithm after refinement of all pixels greater than 1 per cent of the maximum to a resolution of 32". Subsequent panels show refinement to scales of 16", 8", 4" and 2". All panels have been convolved with the synthetic beam of the telescope, and all levels were processed with $\lambda = 300$. }
\label{fig:refinement}
\end{figure}

A step further than this is to solve the system on an adaptive grid, such as a quad-tree. In this case, the system is solved at a low resolution, and pixels where emission is detected are then subdivided. This process continues until the resolution limit of the telescope is reached. As a divide-and-conquer method, this approach reduces the algorithmic complexity of the algorithm, but a detailed investigation of its convergence properties are necessary. To demonstrate its feasibility, we perform a deconvolution using an adaptive quad-tree strategy on a single channel of the NGC2403 dataset used in Section \ref{sec:ngc2403} and the results are shown in Fig. \ref{fig:refinement}. Processing time was around 7 minutes for the channel, a speed up around a factor of 13 compared to solving the system on the complete 1024x1024 grid.

\section{Comparison with existing methods} \label{sec:compare}

\subsection{CLEAN based methods}

The CLEAN algorithm and its variants have been the primary deconvolution methods for radio interferometric imaging for over 40 years. As such, they are extremely mature algorithms and there is a great deal of experience in their use in the community. 

The basic concept behind the CLEAN algorithm is that the image is modelled as a collection of point sources that are built up through an iterative greedy algorithm. This algorithm selects a new point source to be added to the model by determining the residual "dirty image" and selecting the maximum of this image as the location of the next candidate source. The dirty image is calculated from the residual visibilities through the use of the Fourier Transform. 
More recently, an extension to CLEAN to account for the non-coplanar baselines effect has been developed called W-projection \cite{Cornwell:2005p1977}. This method uses a convolution kernel to project the calibrated visibilities to the $w=0$ Fourier plane taking into account the blurring caused by the non-zero $w$ term. Similarly, in the case of direction dependent gains, A-projection kernels were developed by \cite{Bhatnagar:2005p1990} to account for the antenna primary beam patterns in visibility space before the Fourier transform is applied.

If we denote the calibration operation as ${\bf C}$, the Fourier transform as ${\cal F}$, the gridding operation as ${\bf G}$, the model image as ${\bf I}$,  the visibilities as ${\bf V}$, and the dirty image as ${\bf D}$, then the simplest update step of CLEAN can be written as 
\begin{eqnarray}
{\bf D} &  =  & {\cal F}^{-1} \left( {\bf G}{\bf C} {\bf V} - {\cal F}{\bf I} \right)  \\
{\bf I}' & \rightarrow & {\bf I} + \gamma \, {\rm max}{\bf D} \, \delta({\rm arg max}{\bf D})
\label{eq:di}
\end{eqnarray}
where $\gamma$ is the gain of the CLEAN algorithm, and $\delta$ represents a Kronecker delta.

To contrast this to the SL1M algorithm, one can approximate the update step of the algorithm as
\begin{equation}
{\bf I}' \rightarrow {\cal T}\left( {\bf I} + \frac{1}{L} {\bf M}^{-1}\left( {\bf C} {\bf V} - {\bf M} {\bf I}\right) \right).
\label{eq:slim}
\end{equation}
Clearly there are some similarities to the structure of the two algorithms. In particular, there is an analog to the dirty image in SL1M which is calculated through ${\bf M}^{-1}\left( {\bf C} {\bf V} - {\bf M} {\bf I}\right)$. This pseudo dirty image could be used directly in a CLEAN style update step which would update only a single component of the model image, but instead the FISTA $L_1$ minimisation step is used to update all of the image components in a single step.

Recently, \cite{2012ApJ...759...17S} introduced the Fast Holographic Deconvolution method which was used to deconvolve an image created by the MWA 32 antenna prototype. For this CLEAN style algorithm, the update step can be written (loosely) as
\begin{eqnarray}
{\bf D} &  =  & {\cal F}^{-1}{\bf G}{\bf V} -  {\cal F}^{-1} {\bf H}{\cal F}{\bf I}   \\
{\bf I}' & \rightarrow & {\bf I} + \gamma \, {\rm max}{\bf D} \, \delta({\rm arg max}{\bf D})
\label{eq:fhdc}
\end{eqnarray}
where ${\bf G}$ now incorporates the projection effects due to the antenna beams and ${\bf H}$, the holographic mapping function, is introduced. This function distributes the Fourier components of the model image to their correct locations, taking into account the direction dependent gains of the antennas. This Holographic mapping function can be related to the SL1M algorithm by making the identification ${\cal F}^{-1} {\bf H}{\cal F} \rightarrow {\bf M}^{-1}{\bf M}$. Sullivan et al. pre-calculate ${\bf H}$ and store it as a sparse matrix - though they note that this may not be possible when the non-coplanar baseline effect becomes important. This is in contrast to the SL1M algorithm where ${\bf M}$ is dense and calculated in place.

\subsection{Compressive sampling} 

As mentioned earlier, the approach adopted for SL1M parallels closely the approach used in \cite{Li:2011p1778} and \cite{Wenger:2010p2327}. The same basic equations are being solved and the same or a similar L1 minimisation scheme, based on iterative shrinkage and thresholding, is used to solve them. If the update step for Li et al. is written in the same style as above, one has that
\begin{equation}
{\bf I}' \rightarrow  {\cal T}\left({\bf I} + \frac{1}{L} {\cal F}^{-1}\left({\bf G}{\bf C} {\bf V} - {\cal F} {\bf I}\right)\right)
\label{eq:li}
\end{equation}
The fundamental different in the approach in equation (\ref{eq:li}) and SL1M in equation (\ref{eq:slim}) is that the minimisation is done with respect to the calibrated visibilities using the general matrix ${\bf M}$, not on visibilities gridded onto a Fourier plane. The matrix ${\bf M}$ allows a more flexible representation of the relationship between the observed visibilities and the image pixels, at the cost of significantly more computation.

\subsection{Bayesian compressive sensing}

The minimisation problem solved in SL1M, given by equation (\ref{eq:cme}), can be reinterpreted as a maximum a posteriori (MAP) estimate of the reconstructed image given the data.  In this interpretation, the regularisation term represents the prior expectation of the distribution of the reconstructed image values. In this case this prior distribution is an exponential distribution, given by
\begin{equation}
p({\bf I} | \lambda) = \frac{\lambda}{2} \exp\left({-\frac{\lambda}{2} \sum_k\left| I_k \right|}\right).
\end{equation}
Given this interpretation, the shaped of the posterior distribution around the MAP estimate can be explored to determine the errors in the derived image. Approaches such the iterative hierarchical algorithm for solving sparse Bayesian problems as outlined in \cite{Babacan:2010p3398} may be used. Furthermore, this approach includes a method of estimating the covariance of the MAP solution which could used to develop a parameter free algorithm for inverting radio synthesis images, as the noise in the measurements and the sparsity of the solution (represented by $\lambda$) may be inferred from the data using these techniques.

\section{Conclusions}

In this paper we present a new algorithm for deconvolving radio synthesis images based on direct inversion of the measured visibilities that can deal with the non-coplanar base line effect and can be applied to telescopes with direction dependent gains. We have outlined the basic method of the algorithm and demonstrated its application to several synthetic and real datasets showing good reconstruction performance. 

While this algorithm is more computationally demanding than existing methods, it is highly parallelisable and will scale well to clusters of CPUs and GPUs. This algorithm is also extremely flexible, allowing the solution of the deconvolution problem on arbitrarily placed pixels.

More development and investigation of this method is required for its use in solving real-world problems. However, there are many interesting and potentially valuable avenues of investigation. Firstly, the method must be rigorously benchmarked against existing CLEAN implementations for both accuracy and speed and to understand the effect of the regularisation parameter $\lambda$ on the deconvolution result in more detail. Also, minimisation methods other than FISTA should be investigated for faster convergence properties. Secondly, the method should be applied to data from telescopes with direction dependent gains to verify that its performance remains good in this case.  Thirdly, including other established features of radio synthesis software such as multi-scale deconvolution, multi-frequency synthesis and also the inclusion of self-calibration should also be investigated. Finally, deconvolution directly on the HEALPIX grid should be demonstrated, as this is likely to be a valuable feature for future all-sky astrophysics research.

\footnotesize{
\bibliographystyle{aa}
\bibliography{bibfile}

\begin{thebibliography}{21}
\expandafter\ifx\csname natexlab\endcsname\relax\def\natexlab#1{#1}\fi

\bibitem[{Babacan {et~al.}(2010)Babacan, Molina, \&
  Katsaggelos}]{Babacan:2010p3398}
Babacan, S., Molina, R., \& Katsaggelos, A. 2010, IEEE Transactions on Image
  Processing, 19, 53

\bibitem[{Bach {et~al.}(2011)Bach, Jenatton, Mairal, \&
  Obozinski}]{Bach:2011p3079}
Bach, F., Jenatton, R., Mairal, J., \& Obozinski, G. 2011, Foundations and
  Trends in Machine Learning, 4, 1

\bibitem[{Beck \& Teboulle(2009{\natexlab{a}})}]{Beck:2009p2984}
Beck, A. \& Teboulle, M. 2009{\natexlab{a}}, SIAM Journal on Imaging Sciences,
  2, 183

\bibitem[{Beck \& Teboulle(2009{\natexlab{b}})}]{Beck:2009p3004}
Beck, A. \& Teboulle, M. 2009{\natexlab{b}}, IEEE Transactions on Image
  Processing, 18, 2419

\bibitem[{Becker {et~al.}(2011)Becker, Bobin, \& Cand{\`e}s}]{Becker:2011p3142}
Becker, S., Bobin, J., \& Cand{\`e}s, E. 2011, SIAM Journal on Imaging
  Sciences, 4, 1

\bibitem[{Bhatnagar {et~al.}(2005)Bhatnagar, Golap, \&
  Cornwell}]{Bhatnagar:2005p1990}
Bhatnagar, S., Golap, K., \& Cornwell, T.~J. 2005, Astronomical Data Analysis
  Software and Systems XIV ASP Conference Series, 347, 96

\bibitem[{Candes \& Romberg(2005)}]{Candes:2005p3330}
Candes, E.~J. \& Romberg, J.~K. 2005, in Electronic Imaging 2005, ed. C.~A.
  Bouman \& E.~L. Miller (SPIE), 76--86

\bibitem[{Carrillo {et~al.}(2012)Carrillo, McEwen, \& Wiaux}]{Carrillo:2012ho}
Carrillo, R.~E., McEwen, J.~D., \& Wiaux, Y. 2012, Monthly Notices of the Royal
  Astronomical Society, 426, 1223

\bibitem[{Cornwell(2008)}]{Cornwell:2008p1783}
Cornwell, T.~J. 2008, arXiv, astro-ph, 793

\bibitem[{Cornwell {et~al.}(2005)Cornwell, Golap, \&
  Bhatnagar}]{Cornwell:2005p1977}
Cornwell, T.~J., Golap, K., \& Bhatnagar, S. 2005, Astronomical Data Analysis
  Software and Systems XIV ASP Conference Series, 347, 86

\bibitem[{Cornwell {et~al.}(2012)Cornwell, Voronkov, \&
  Humphreys}]{Cornwell:2012p1785}
Cornwell, T.~J., Voronkov, M.~A., \& Humphreys, B. 2012, Proc. SPIE, 8500 Image
  Reconstruction from Incomplete Data VII, 85000L

\bibitem[{Deboer {et~al.}(2009)Deboer, Gough, Bunton, Cornwell, Beresford,
  Johnston, Feain, Schinckel, Jackson, Kesteven, Chippendale, Hampson,
  O'Sullivan, Hay, Jacka, Sweetnam, Storey, Ball, \& Boyle}]{Deboer:2009p3117}
Deboer, D.~R., Gough, R.~G., Bunton, J.~D., {et~al.} 2009, Proceedings of the
  IEEE, 97, 1507

\bibitem[{H{\"o}gbom(1974)}]{Hogbom:1974p2738}
H{\"o}gbom, J.~A. 1974, Astronomy and Astrophysics Supplement, 15, 417

\bibitem[{Li {et~al.}(2011)Li, Cornwell, \& Hoog}]{Li:2011p1778}
Li, F., Cornwell, T.~J., \& Hoog, F.~d. 2011, Astronomy and Astrophysics, 528,
  A31

\bibitem[{Marsh \& Richardson(1987)}]{Marsh:1987uc}
Marsh, K.~A. \& Richardson, J.~M. 1987, Astronomy and Astrophysics, 182, 174

\bibitem[{McEwen \& Wiaux(2011)}]{McEwen:2011p3034}
McEwen, J.~D. \& Wiaux, Y. 2011, Monthly Notices of the Royal Astronomical
  Society, 413, 1318

\bibitem[{Narayan \& Nityananda(1986)}]{Narayan:1986p3297}
Narayan, R. \& Nityananda, R. 1986, IN: Annual review of astronomy and
  astrophysics. Volume 24 (A87-26730 10-90). Palo Alto, 24, 127

\bibitem[{Sullivan {et~al.}(2012)Sullivan, Morales, Hazelton, Arcus, Barnes,
  Bernardi, Briggs, Bowman, Bunton, Cappallo, Corey, Deshpande, deSouza,
  Emrich, Gaensler, Goeke, Greenhill, Herne, Hewitt, Johnston-Hollitt, Kaplan,
  Kasper, Kincaid, Koenig, Kratzenberg, Lonsdale, Lynch, McWhirter, Mitchell,
  Morgan, Oberoi, Ord, Pathikulangara, Prabu, Remillard, Rogers, Roshi, Salah,
  Sault, Udaya~Shankar, Srivani, Stevens, Subrahmanyan, Tingay, Wayth,
  Waterson, Webster, Whitney, Williams, Williams, \&
  Wyithe}]{2012ApJ...759...17S}
Sullivan, I.~S., Morales, M.~F., Hazelton, B.~J., {et~al.} 2012, The
  Astrophysical Journal, 759, 17

\bibitem[{Wenger {et~al.}(2010)Wenger, Magnor, Pihlstr{\"o}m, Bhatnagar, \&
  Rau}]{Wenger:2010p2327}
Wenger, S., Magnor, M., Pihlstr{\"o}m, Y., Bhatnagar, S., \& Rau, U. 2010,
  Publications of the Astronomical Society of the Pacific, 122, 1367

\bibitem[{Wiaux {et~al.}(2009{\natexlab{a}})Wiaux, Jacques, Puy, Scaife, \&
  Vandergheynst}]{Wiaux:2009p3076}
Wiaux, Y., Jacques, L., Puy, G., Scaife, A. M.~M., \& Vandergheynst, P.
  2009{\natexlab{a}}, Monthly Notices of the Royal Astronomical Society, 395,
  1733

\bibitem[{Wiaux {et~al.}(2009{\natexlab{b}})Wiaux, Puy, Boursier, \&
  Vandergheynst}]{Wiaux:2009p3050}
Wiaux, Y., Puy, G., Boursier, Y., \& Vandergheynst, P. 2009{\natexlab{b}},
  Monthly Notices of the Royal Astronomical Society, 400, 1029

\end{thebibliography}
}
\end{document}